\documentclass[12pt]{article}

\usepackage{authblk}

\usepackage{geometry}
\usepackage[parfill]{parskip}

\usepackage{amsmath, amssymb, amsthm}

\usepackage{graphicx}

\usepackage{hyperref}

\usepackage{caption}
\usepackage{subcaption}

\usepackage{xcolor}
\usepackage{fancyvrb}

\usepackage{tikz}
\usetikzlibrary{shapes.geometric, arrows.meta}

\renewcommand{\thefootnote}{\fnsymbol{footnote}}

\title{Visualising spatio-temporal health data: the importance of capturing
the 4th dimension
}

\author[1]{Alison C. Hale \thanks{Corresponding author: Alison Hale, a.c.hale@lancaster.ac.uk }}
\author[1]{Charlotte Appleton}
\author[2]{P.-J. M. Noble}
\author[2]{Gina L. Pinchbeck}
\author[1]{Barry Rowlingson}
\author[1]{Peter J. Diggle}
\author[2]{Alan D. Radford}
\author[1]{Christopher P. Jewell}

\affil[1]{Lancaster Medical School, Lancaster University, Lancaster, UK. LA1 4YW}
\affil[2]{Institute of Infection, Veterinary and Ecological Sciences, University of Liverpool, Leahurst Campus, Neston, UK. CH64 7TE}

\begin{document}

\maketitle
\renewcommand{\thefootnote}{\fnsymbol{footnote}}

\clearpage
\newpage

\begin{center}
\textbf{Abstract}
\end{center}

\leftskip1.5cm
\rightskip1.5cm

    Confronted by a rapidly evolving health threat, such as an infectious disease outbreak, it is essential that decision-makers are able to comprehend the complex dynamics not just in space but also in the 4th dimension, time.
    In this paper this is addressed by a novel visualisation tool, referred to as the Dynamic Health Atlas web app, which is designed specifically for displaying the spatial evolution of data over time while simultaneously acknowledging its uncertainty.
    It is an interactive and open-source web app, coded predominantly in JavaScript, in which the geospatial and temporal data are displayed side-by-side.
    The first of two case studies of this visualisation tool relates to an outbreak of canine gastroenteric disease in the United Kingdom, where many veterinary practices experienced an unusually high case incidence.
    The second study concerns the predicted COVID-19 reproduction number along with incidence and prevalence forecasts in each local authority district in the United Kingdom.
    These studies demonstrate the effectiveness of the Dynamic Health Atlas web app at conveying geospatial and temporal dynamics along with their corresponding uncertainties.

\leftskip0cm
\rightskip0cm

\bigskip
\bigskip
\bigskip

\section{Introduction}
    Over the last couple of decades there has been an ongoing explosion in the collection, analysis and interpretation of data.
    This is made possible by increasingly sophisticated technologies along with corresponding advances in data mining and computer modelling.
    This explosion is being driven by the fact that data is both ever more plentiful and pervasive: 
    the International Data Corporation (IDC) estimated that the Global Datasphere was 33 Zettabytes (ZB) in 2018, rising to 59ZB by 2020, and forecast to reach 175ZB by 2025 \cite{Reinsel2018}.
    Harnessing this information has the potential to aid decision making at all levels, from international to personal, and as such could potentially benefit everyone.
    In the context of human and veterinary healthcare, the intelligent use of information and technology provides decision-makers with opportunities to better understand a population's state of health and thereby provide improvements to healthcare along with the possibility of instigating preemptive actions \cite{NatMed2020, Tomines2013}.

    A visually powerful and informative way to represent population-level health data is to aggregate it over space and time, and where applicable it can be mapped onto a geographic region.
    There are numerous examples of such maps where the health data is varying relatively slowly over time including: Map Gallery by the World Health Organisation \cite{WHOmaps}; Global Burden of Disease (GBD) Data Visualizations by the Institute for Health Metrics and Evaluation \cite{IHMEmaps}; The Environment and Health Atlas for England and Wales by the UK Small Area Health Statistics Unit (SAHSU) \cite{SAHSUmaps}.
    Although less common there are also health maps where the data is rapidly changing, usually in the context of infectious disease outbreaks, for instance: HealthMap utilises online sources for disease outbreak monitoring and real-time surveillance of emerging public health threats \cite{HealthMapBCH}; COVID-19 Dashboard monitors global cases and deaths \cite{covidCSSEJHU}; COVID-19 Surveillance Dashboard monitors global cases and vaccinations \cite{covidBIUV}.
    Health maps, including the aforementioned, are typically static in that the user cannot scroll through time to see a sequence of maps changing over time, and furthermore in some cases there is no, or a limited, facility to examine temporal trends e.g. on a time series.
    With these health maps it is difficult, sometimes impossible, for the intended end-user(s) to fully grasp both the spatial and temporal dynamics of the data, yet this is essential for understanding how a disease threat is evolving.
    This lack of understanding inhibits, potentially significantly, the ability of policy-makers to make informed decisions along with proportionate interventions.

    To simultaneously display the evolution over time of geospatial data on a surface along with its corresponding uncertainties requires a visualisation in 4 dimensions  i.e. time, latitude, longitude, data value.
    Note that in this sense data values along with their corresponding uncertainty are displayed in a dimension orthogonal to the geospatial surface and time.
    The uncertainties acknowledge the error in the data whether it was measured or predicted from a model.
    Displaying geospatial data on a surface is generally achieved by colour coding data points/regions on a map, critically this fails to acknowledge the uncertainties in the data and also its temporal evolution.
    Conceivably the temporal evolution over a geospatial surface could be visualised by creating a movie where each frame represents a time slice, however this would fail to acknowledge uncertainties.
    
    In this paper an approach is presented for addressing these issues of visualising across 4 dimensions.
    To this end this work describes the development of the Dynamic Health Atlas (DHA), an interactive web app specifically designed to provide dynamic and interactive visualisations of geospatial-temporal data.
    This web app displays data values mapped over a geographic region and a time series is used to convey the temporal evolution of each geographic sub-region/point of interest.
    The time series also displays the uncertainty relating to the data.
    The user is able to interact with the web app allowing them to select any spatial and temporal slices of interest.
    Apart from being widely accessible a web app has the added advantages that it can be hosted on any web server and it is potentially straightforward to frequently update its data source.
    Furthermore since the DHA is standalone visualisation software it is completely separate from the process of modelling or measuring the geospatial-temporal data.
    In this paper the DHA is applied to two case studies in the United Kingdom both of which required daily updates: a) predictions relating to data collected at veterinary practices during an outbreak of gastrointestinal disease in companion dogs; b) forecasts of COVID-19 cases in local authority districts.

\section{Methods: Software development}

\subsection{Dynamic Health Atlas (DHA)}
    
    The overall objective underpinning the DHA was to develop a generic reusable standalone web app for displaying geographically referenced population-level health data. 
    A core requirement was that it should be able to display geospatial and temporal data side-by-side while also conveying the uncertainty in this data: example screenshots are given in Figures \ref{fig:caninegastro} and \ref{fig:covid}.
    This web app's interface was designed around the needs of the end-user.
    The UI/UX (user interface/user experience) were a central design requirement in that a policy-maker or health professional should find it straightforward and intuitive to use thereby allowing them to focus on understanding and interpreting the data.
    Accordingly a balance needed to be found between presenting too little and too much information while simultaneously giving the end-user enough control to select different slices of data: for example select a time slice to be displayed on the map or select a time series plot for a given geographic location.
    
\begin{figure}[ht!]
    \centering
    \begin{tikzpicture}[node distance=2cm]
    
    \tikzstyle{startstop} = [rectangle, rounded corners, minimum width=3cm, minimum height=1cm,text centered, draw=black, fill=red!20]
    \tikzstyle{io} = [trapezium, trapezium left angle=70, trapezium right angle=110, minimum width=3cm, minimum height=1cm, text centered, draw=black, fill=blue!20]
    \tikzstyle{process} = [rectangle, minimum width=3cm, minimum height=1cm, text centered, draw=black, fill=green!20]
    \tikzstyle{decision} = [diamond, minimum width=3cm, minimum height=1cm, text centered, draw=black, fill=orange!20]
    \tikzstyle{arrow} = [thick,-{Stealth[scale=1.5]}]

    \node (start) [startstop] {space-time data};
    \node (pro1)  [process, below of=start] {ETL};
    \node (in1)   [io, below of=pro1, xshift=-2.5cm] {GeoJSON};
    \node (in2)   [io, right of=in1, xshift=3cm] {config. file};
    \node (pro2)  [process, below of=in1, xshift=2.5cm] {DHA};
    \node (stop)  [startstop, below of=pro2] {web browser};

    \draw [arrow] (start) -- (pro1);
    \draw [arrow] (pro1) -- (in1);
    \draw [arrow] (pro1) -- (in2);
    \draw [arrow] (in1) -- (pro2);
    \draw [arrow] (in2) -- (pro2);
    \draw [arrow] (pro2) -- (stop);

    \end{tikzpicture}
    \caption{Outline of data setup workflow needed for prepping the data which is read by the DHA.
    The GeoJSON and configuration files are created offline.
    The input `space-time data' refers to unprocessed geospatial-temporal data. Note that ETL denotes extract-transform-load.}
    \label{fig:DHAflowChart}
    \end{figure}
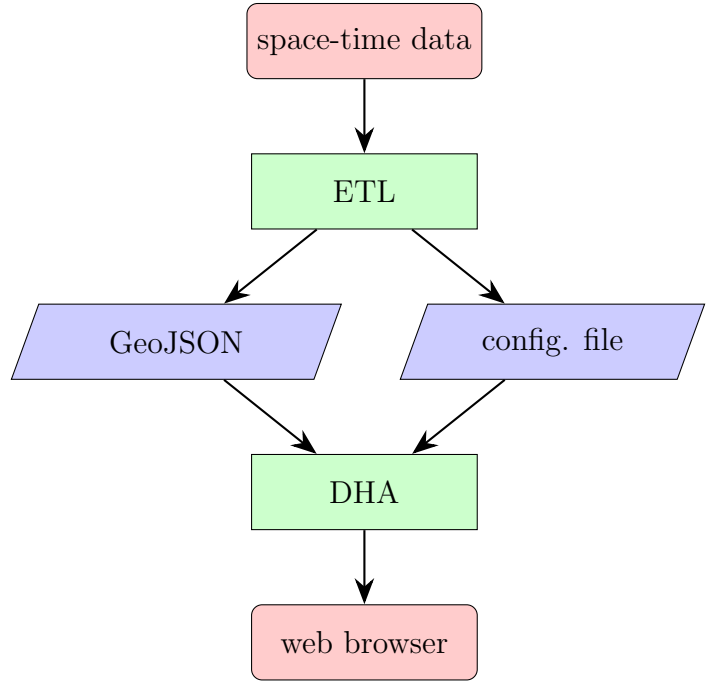

    The raw geospatial-temporal health data needs to be transformed such that it can be displayed using the DHA, the overall process used to achieve this is shown in Figure \ref{fig:DHAflowChart}.
    The DHA's main software engineering objectives were to make the web app open source and lightweight.
    The former would allow developers to freely reuse and augment the web app in the future.
    The latter relates to keeping the web app's use of system resources, specifically browser RAM and network communications, to a minimum so that it would be compatible with a variety of networks (e.g. mobile).
    There are a number of viable software architecture routes that could been taken, including server-side frameworks such as the Dash \cite{dashPiFr} and Shiny \cite{Shiny} frameworks: alternative architectures/frameworks are considered in the Discussion Section.
    Here the aforementioned software objectives underpinning the DHA are achieved in a client-side implementation using the JavaScript programming language.
    To this end two lightweight open source JavaScript libraries are used: Leaflet \cite{Leafletjs2018} for mapping spatial data on a geographic region; and Chart.js \cite{Chartjs2018} for plotting time series data.
    These two libraries were not designed to work together so a substantial quantity of JavaScript was written to integrate them into the DHA.
    The main role of this JavaScript was to manage spatial and temporal data such that each location on the map correctly corresponded with its temporal data and vice versa.

    For visual impact the web app was designed to display everything on a single web page constructed using HTML5 and CSS3.
    For example standard HTML input tags, drop-down lists and range sliders, were chosen as the means to select the required slice of data.
    This web page was designed to have two layouts: `landscape' intended for standard laptop/computer screen (width~$>$~1280px) and `portrait' intended for tablet screens especially in portrait orientation (width~$\sim$~860px).
    Widths of less that 860px are not accommodated as they were considered too small to effectively display the data side-by-side.

    The JavaScript library Leaflet reads data using the open standard GeoJSON format \cite{Butler2016}.
    This format, based on the JSON format, represents simple geographical features along with their corresponding attributes.
    Whereas the JavaScript library Chart.js reads time series data using the JSON format.
    The traditional GIS (geographic information system) formats, such as standard GeoJSON, contain data value(s) for each geographic feature.
    Whereas time series data is better represented by the JSON format, that is by using a key:value pair where each `key' codes a time point.
    In terms of the DHA it would be inefficient to store the geospatial-temporal data separately using both the GeoJSON and JSON formats.
    To avoid this duplication the design decision was made to store both types of data in the GeoJSON format.
    
    The GeoJSON format contains a Feature object which represents either a spatially bound surface or a spatial point.  Each Feature object should contain: a ``type'' member with value ``Feature''; a ``geometry'' member whose value is a Geometry object representing points/surfaces in coordinate space; and a ``properties'' member whose value is any JSON object.
    The ``properties'' member is therefore a flexible object.
    
    For geospatial data alone the value of the ``properties'' member is typically a single key:value pair where the value is displayed on a map.
    However the DHA requires geospatial-temporal data so the value of the ``properties'' member is a JSON object containing all the temporal data as a list of key:value pairs.
    For instance this list takes the form ``0'':12, ``1'':15, ``2'':10, etc., where the `key' string codes time and `value' is the corresponding health data value, these values are displayed on both the map and time series e.g. median prevalence at time ``0''.
    An example of is shown in the Appendix.
    In this adapted-GeoJSON format time was always encoded using sequential integers starting from zero (i.e. first time point in time series).
    The uncertainty values are similarly stored in the value of the ``properties'' member where the JSON object `key' from the key:value encodes both the time point and its corresponding uncertainty level e.g. lower or upper quantile: these values are displayed on the time series.
    For instance the key:value pairs ``0.L'':7 and ``0.U'':19 refer to the lower and upper quantiles respectively for the first time point in the time series, the second time point key:value pairs are ``1.L'':9 and ``1.U'':22, and so on.
    Note that ``.L'' and ``.U'' are used for whatever measure of uncertainty is appropriate for the given data e.g. quantiles, confidence intervals, credible intervals, etc.

    In terms of the DHA, for each feature the time series uses the JSON object from the ``properties'' member in order to display the temporal values of the health data (e.g. median) along with its corresponding uncertainties.
    The map displays, simultaneously for all features, the health data (e.g. median) for the currently selected time-slice.
    To this end the DHA must keep track of which temporal layer is displayed on the map and which feature's temporal values are depicted on the time series.

    The layout and appearance of the maps and time series on the DHA are defined using a configuration file.
    In particular this allows a developer to configure the colours on the maps, legend intervals, time series dates and so on.
    This configuration file also contains initial values including: coordinates to center the map; map zoom level; and map distance units (i.e. miles or kilometers).
    Note that the configuration file can be created or edited without any need to edit the JavaScript code underlying the DHA.

    The DHA is open source with a MIT licence and can be configured and deployed using any suitable dataset.
    A tutorial explaining how to configure the web app is available at \url{https://achale.gitlab.io/dynamicatlastutorial/} \cite{DHAtutorial}.
    The DHA source code is available from GitLab at \url{https://gitlab.com/achale/dynamicatlas/} \cite{DHAcode}.
    A corresponding demo is included which can be viewed at \url{https://achale.gitlab.io/dynamicatlas/} \cite{DHAdemo}, where GitLab Pages was used to create a website directly from the repository.
    The DHA depends on other open source JavaScript libraries including Leaflet and Chart.js, therefore to ensure the current version of the DHA on GitLab is reliable the version numbers of its dependencies are locked.

\subsection{Configuration app using Shiny}
\label{sec:ShinyWebApp}
    Shiny \cite{Shiny, shinyForR} is a package for building interactive web apps using the R programming language \cite{Rcran2021}, it can be used for creating Shiny apps on a web page.
    A configuration app using Shiny was developed and deployed on a Shiny server to ease the process of creating the adapted-GeoJSON and configuration files for the DHA.
    With reference to Figure \ref{fig:DHAflowChart} this configuration app is an example ETL (extract-transform-load) process, note that in principle any appropriate ETL process can be used to generate the configuration and adapted-GeoJSON files.
    
    The configuration app was designed to allow a developer, at their discretion, to upload a suitable set of health data (format: CSV file) along with its corresponding geographic data (format: shapefiles \cite{ESRI1998}) to the host Shiny server.
    The data is not stored on the server beyond the web browser session.
    This configuration app provides a web form which is used to configure the layout and appearance of the map and time series on the DHA web app.
    The web form contains a key:value pair along with a description for each setting (e.g. map legend intervals, time series axis labels, colouring of data on map).
    Using this information along with the uploaded data the configuration app generates the adapted-GeoJSON, corresponding configuration file and automatically loads everything into an instance of the DHA: this allows the developer to check how their data will appear on the DHA.
    Note that this particular version of the DHA is embedded within the configuration app on the Shiny server to allow the two apps to seamlessly communicate.
    The adapted-GeoJSON and configuration files can be downloaded and if needed the user can move back and forth between the configuration app and DHA to refine the layout and appearance of the maps and time series.
    As a result the configuration app allows researchers and developers to create their own maps without needing an expert knowledge of web development.
    
    The configuration app can be accessed at
    \url{http://fhm-chicas-apps.lancs.ac.uk/shiny/users/haleac/healthatlas/} \cite{ShinyDHA}.

\section{Case studies}
    Two case studies are presented in which the DHA is used to display predictions.
    The first relates to predicting an outbreak of gastroenteric disease in companion dogs during the winter of 2019/20.
    The second relates to forecasting COVID-19 cases during the first part of 2021, one year into the global pandemic.
    
    Both case studies discussed required their own pipeline from the source data to `outcome' data i.e. data displayed by the DHA.
    In both studies the source data and the outcome data were updated each day, so the DHA always displayed the most recent data.
    The pipeline for each study had three main stages:
    \begin{enumerate}
        \item
        The `static' and `dynamic' source data were collected and preprocessed into a suitable format for the next stage in the pipeline.
        The `static' source data included geographic information (latitude and longitude coordinates of geo-coded data) and one or more explanatory variables e.g. index of multiple deprivation, commuting patterns, etc.
        The `dynamic' source data included the daily case incidence of disease at each location along with one or more explanatory variables e.g. age, sex, etc.
        
        \item 
        Given the preprocessed source data a spatio-temporal prediction model was fitted using Bayesian inference.
        In particular a Markov Chain Monte Carlo (MCMC) algorithm was used to draw samples from the posterior distribution.
        Each of the case studies used its own prediction model which, using parallel processing, ran in a few hours overnight.

        \item
        Summary statistics at each location and time were calculated from the posterior samples.
        With reference to Figure \ref{fig:DHAflowChart} these summary statistics are the `space-time data'.
        In other words these statistics along with their corresponding geographic data were used 
        as the source for an ETL process, the output of which was the adapted-GeoJSON and configuration files. 
        Note that here the ETL process was written in a programming language so that it could be run as part of a shell script, hence it does not use the configuration app described in Section \ref{sec:ShinyWebApp}.
        The adapted-GeoJSON and configuration files were transferred to, and made live on, the web server where the DHA was hosted.
        In the case of the COVID-19 case study a GitLab CI/CD process was configured to automatically transfer updated files to the repository's web front-end i.e. GitLab Pages.
        
    \end{enumerate}
    For quality assurance reasons each pipeline was semi-automated but not fully automated.
    Specifically the posterior samples were checked, by human eyes, each time a prediction model was run with new source data.
    To this end MCMC diagnostics included: trace plots, density plots and autocorrelation plots.
    Similarly the DHA maps along with their corresponding time series' were also examined before they were made live on the web server.
    
    Before describing the case studies in detail a general overview of the DHA page layout, as shown in Figures~\ref{fig:caninegastro} and \ref{fig:covid}, is detailed here:
    \begin{itemize}
        \item The drop-down list (top left) is used for selecting the type of data to be displayed by the map and time series.
        \item Coloured circles or regions on the map display the geospatial data. The colour coding intervals are shown on the legend located within the map frame (bottom right).
        \item When a given coloured circle or region is selected, e.g. by clicking on it, its corresponding pointwise temporal data is displayed on a time series (top right).
        \item The position of the vertical red line on the time series denotes which time slice of the geospatial data is shown on the map.
        \item The time slice of the geospatial data displayed on the map can be changed to any time point on the time series x-axis by using either the range slider or red arrows located under the time series. One click on a red arrow increments one time unit.
        Clicking once on a blue arrow (either side of `play') sequentially increments through all time units, hence time slices of the geospatial data are shown on the map one after another: the speed at which it increments is either the default speed or it is set by using the `select speed' drop-down list.
        \item The four blue icons in the top right corner of the web page have the following functions: far-left, link to an information page about the data displayed on the maps; middle-left, link to download the data being displayed on the map; middle-right, toggle web page between landscape and portrait layout; far-right, link to a help page describing how to interact with the web app.
    \end{itemize}

\subsection{Canine gastroenteric disease predictions}

    Anonymised electronic health records (EHRs) for companion animals were collected by the Small Animal Veterinary Surveillance Network (SAVSNET), a volunteer network of 301 veterinary practices (663 premises) in the United Kingdom \cite{Fernando2015, Fernando2017}.
    Between 29 February 2014 and 17 March 2020 a total of 7,094,397 EHRs were collected, of which 4,685,732 related to dogs.
    During each veterinary consultation the data collected in the EHRs included: age, breed, neuter status, owners' postcode, sex, species and vaccination status.
    Additionally at the time of the consultation it was compulsory for the veterinary clinician to record one main presenting complaint from the following: gastroenteric, respiratory, pruritus, tumor, kidney disease, other unwell, post-op check, vaccination or other healthy.
    Gastroenteric disease is known to be a common complaint among dogs presenting for veterinary care, e.g. \cite{ONeill2014}, however during January 2020 a significant increase was observed by SAVSNET.
    A research study conducted during the first part of 2020 concluded that a potential cause of the  outbreak was a canine enteric coronavirus \cite{Radford2021}.
    Ethics approval was given by Liverpool University Research Ethics Committees (Liverpool, UK; VREC922/RETH000964).

    In this paper detection of the aforementioned outbreak of canine gastroenteric disease was undertaken using a spatio-temporal mixed effects regression model which was fitted separately for each country using the `caramellar' package \cite{caramellar2017}.
    A full description of this model was published previously: `A real-time spatio-temporal syndromic surveillance system with application to small companion animals' \cite{Hale2019}.
    In the following, for convenience, an outline of the modelling approach is reiterated.
    An outbreak, localised in space and time, of a given disease is defined as an unexplained rise in the ratio between consultations recording the disease and all other consultations.
    The mixed effects regression model is composed of two parts: `fixed effects' which in this work are accounted for by a multiple linear regression; and `random effects' which describe a latent stochastic process.
    Variations in the underlying population caused by well understood phenomena are accounted for with measured explanatory variables, these enter into the `fixed effects'. 
    However the population will also be subject to unexplained fluctuations in space and time, these are accounted for by a latent, spatially and temporally correlated stochastic process $S_{i,t}$ at location $i$ and time $t$.
    Theoretically the expected value of $S_{i,t}$ is zero although in practice it will never be precisely zero.
    To this end a user-specified threshold $l$ is defined as a reference to determine if $S_{i,t}$ is materially greater than zero.
    Specifically the probability conditional on all available data up to and including time $t$ is referred to as the predictive probability $q$, the aim therefore is to compute the predictive probability that $S_{i,t}>l$ at each location and time.
    Relative to a given threshold $l$, 
    an outbreak at a given location is declared when the predictive probability $q$ exceeds a prespecified level $q_0$ e.g. $0.9$.
    
    In terms of the results depicted on the DHA in this paper the specifics of the model are as follows.
    Location $i$ refers to a veterinary premise and time unit $t$ refers to either day or week.
    The mixed effects regression model has a binary response variable, it takes the value $1$ for each gastroenteric consultation and $0$ otherwise.
    The fixed effects part of the mixed effects regression model is described by a multiple linear regression with the following explanatory variables: age (years), age squared (years$^2$), sex (female/male), purebred (true/false) and index of multiple deprivation IMD at the owners' address.
    (The most recent IMD data from the Office for National Statistics \cite{ONSgeoPortal2021} was sourced from doogal.co.uk \cite{doogal2020}.)
    Note that here the IMD is a continuous variable computed using a given location's IMD rank divided by its country's maximum rank hence $0<$~IMD~$\leq1$.
    For the purposes of this paper the threshold $l$ is set at $0.1$, this corresponds to $1.25$ times more consultations recording canine gastroenteric disease than would usually be expected on historic grounds: historically the probability of canine gastroenteric disease consultations is 0.03.
    Furthermore a predictive probability $q$ of greater than $q_0=0.9$ is chosen to signal an outbreak, in other words it is highly likely that there are an unusually high proportion of canine gastroenteric consultations.
    When interpreting the model’s results it should  be  noted  that it uses data  based  on  the ratio of counts of canine gastroenteric consultations to all consultations therefore its estimates will be sensitive to fluctuations in the denominator.
    Outbreak predictions using this model ceased at the end of the first week in March 2021 (week 10 of 2020) because there were significant changes in the pattern of consultations caused by the initial stages of the COVID-19 global pandemic.
    Moreover at this time it was also apparent that the canine gastroenteric disease outbreak had considerably subsided.

\begin{figure}[ht!]
    \centering
    \includegraphics[width=0.95\textwidth]{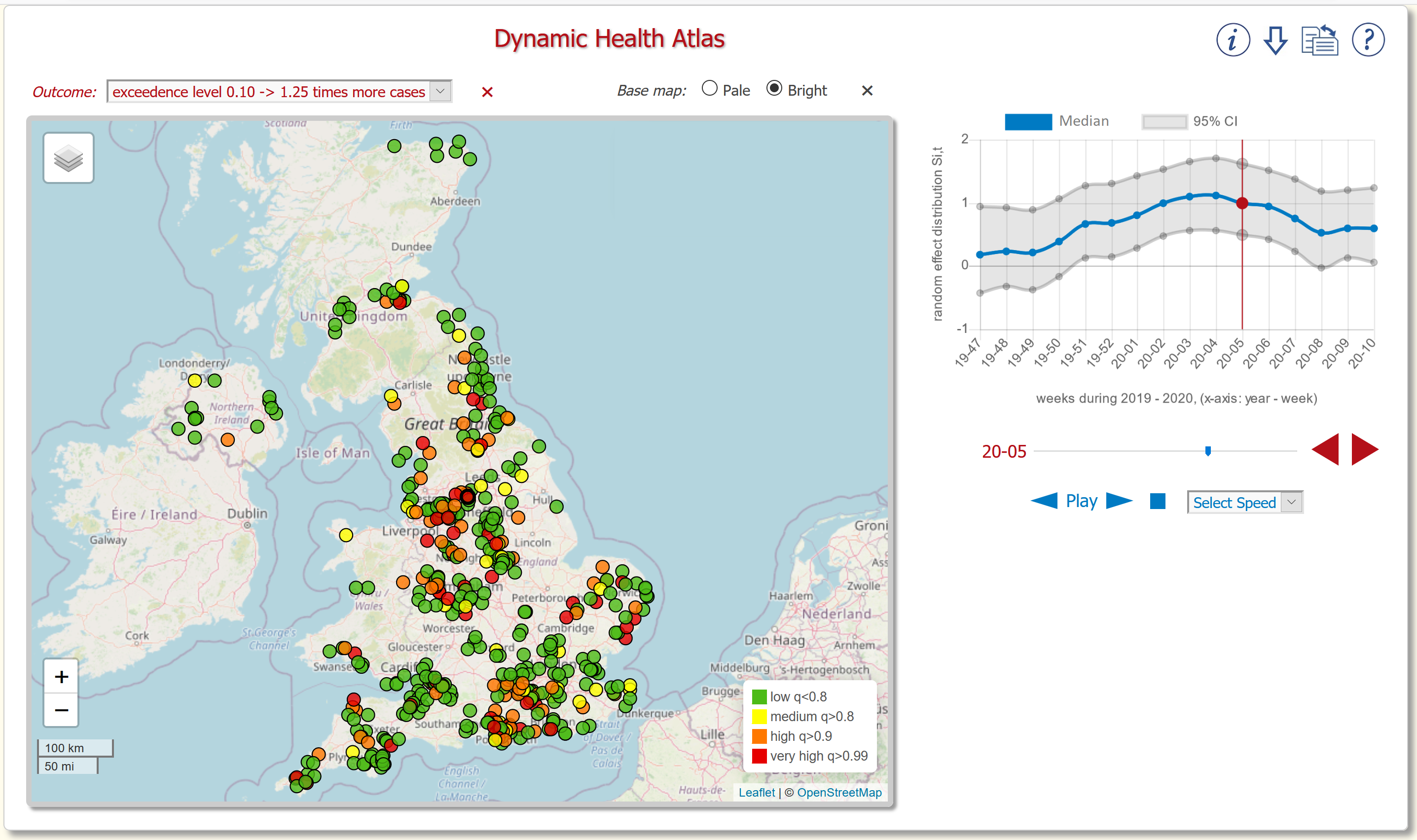}
    \caption{DHA relating to predictions of an outbreak of canine gastroenteric disease recorded at veterinary practices across the UK from week $47$ of 2019 through to week $10$ of 2020.
    The map shows predictions for week $5$ in 2020.
    The time series displays the predictions for a premise between Leeds and Manchester.
    The base map is produced by the OpenStreetMap Foundation using OpenStreetMap data under the Open Database License \cite{OpenStreetMap}.
    }
    \label{fig:caninegastro}
    \end{figure}

    The screenshot of the DHA in Figure~\ref{fig:caninegastro} shows predictions relating to the outbreak of canine gastroenteric disease.
    In this context the overall layout of the web app shown in this figure includes the following features:
    \begin{itemize}
        \item The drop-down list (top left) selects prediction for different thresholds $l$ (exceedence levels), $l=0.1$ in Figure~\ref{fig:caninegastro}.  This was the lowest threshold, other thresholds not shown were $0.19, 0.33, 0.54$.
        \item The coloured circles on the map represent the locations of veterinary premises. When a given premise's coloured circle is selected its corresponding time series shows the dynamics of the disease outbreak at that premise.
        \item The predictions shown on the map are for week $5$ in 2020 as denoted by the vertical red line on the time series.
        The time series shows pointwise predictions, along with their corresponding uncertainties, made at weekly intervals from week 47 of 2019 through to week 10 of 2020.
    \end{itemize}
    The web app is designed to allow any available time slice to be viewed on the map, this enables the end-user to visualise how the epidemic is evolving over time across the whole country.
    Additionally the time series can be viewed at any premise, therefore the end-user can examine the local temporal dynamics.
    As shown in Figure \ref{fig:caninegastro} the epidemic is first increasing and then subsiding in a matter of weeks therefore including time in this manner allows the end-user, e.g. policy-maker, to explore the epidemic's spatial and temporal extent both locally and globally; this would not be possible if the map was static e.g. displayed only the most recent prediction.

    The particulars of the predictions shown in Figure~\ref{fig:caninegastro} are as follows.
    All the predictions in are based on consultation data from participating veterinary practices up to and including week $10$ of 2020.
    The map depicts, at each veterinary premise $i$ for week $5$ of 2020, the predictive probability $q$ that $S_{i,t}>l$ where $l=0.1$.
    Specifically it is colour coded according to the predictive probability $q$ exceeding $q_0=0.99$ (red), $q_0=0.9$ (orange), $q_0=0.8$ (yellow), otherwise $q$ is less than $q_0=0.8$ (green).
    When the predictive probability is high, e.g. $q>0.9$, it is very likely there is an unusually high proportion of gastroenteric consultations.
    Conversely $q<0.8$ indicates the proportion is most likely not high; in practise this suggests that gastroenteric consultations are likely to be around their usual level especially given the relatively low threshold of $l=0.1$.
    As can be seen in Figure~\ref{fig:caninegastro}, during week 5 of 2020, there are a significant number of premises with red and orange circles indicating an unusually high proportion of gastroenteric consultations recorded at these premises.
    
    Each veterinary premise $i$ has its own time series, each plot displays the predicted distribution of $S_{i,t}$ where the blue line is the pointwise median.
    The uncertainty in this prediction is acknowledged by the pointwise $95$\% predictive intervals ($0.025$ to $0.975$ quantiles), this relates to the grey shaded region of the time series.
    For the premise (between Leeds and Manchester) used in Figure~\ref{fig:caninegastro} the time series shows that $S_{i,t}$ starts to rise appreciably towards the final weeks in 2019, then reaches a peak during the forth week of 2020 before subsiding.
    This temporal pattern is generally representative of how the outbreak developed over the United Kingdom.
    The model predictions indicated the outbreak peaked around week 4 and 5 of 2020, this is when there was the largest number of premises with $q>0.9$.
    It should be noted that not all premises saw a peak during these weeks, some peaked earlier, and some later.
    Although not shown there were a total $10$ premises in week 5 of 2020, including the one whose time series is shown in Figure~\ref{fig:caninegastro}, where the predictive probability $q$ exceeded $q_0=0.9$ with $l=0.54$, this indicates three times more consultations of canine gastroenteric disease than would normally be expected at these premises.
    The results discussed above are consistent with the outbreak reported in \cite{Radford2021}.
    
\subsection{COVID-19 forecasts in the United Kingdom}

    In this section we use the UK Health Security Agency (UKHSA) Pillar 2 community testing data is used to identify positive cases of COVID-19 in the United Kingdom.
    This data was collected from SARS-CoV-2 swab testing in the community using either by polymerase chain reaction (PCR) tests or a lateral flow test (LFT): note that cases were not double counted if an individual reported multiple positive test results.
    More details regarding Pillar 2 testing can be found on the UK Government website \cite{DHSCPillars2021} and the data is available at \cite{GOVUKcovid19casedata}.

    The population-level dynamics of COVID-19 epidemic in the United Kingdom during 2020/21 were such that at different times and places localised spikes in disease incidence often caused significant increases of disease in the surrounding area.
    In addition some infectious individuals who travelled seeded new spikes further afield.
    Consequently the epidemic had complex dynamics which were rapidly changing over both time and space.
    In contrast to static maps the DHA is well suited to this spatio-temporal data.
    For reference the evolution of the epidemic over time is shown in Figure \ref{fig:PHEcases} in the Appendix, it depicts the daily case incidence from January 2020 to April 2021.

    Disease transmission is modelled using stochastic compartmental model within a fully Bayesian framework.
    The model assumes that each individual in the population is assigned to one of a number of states, over time individuals transition between states.
    In our model the four states are: susceptible (S), exposed but not yet infectious (E), infected/infectious (I) and recovered/died (R).
    Over time individuals transition sequentially through the states as follow: S$\rightarrow$E, E$\rightarrow$I and I$\rightarrow$R.
    To assist policy-makers with their decision making the posterior samples from the fitted model were used to estimate a number of metrics including reproduction numbers along with forecasts of case incidence and prevalence: all metrics were estimated on a daily basis for each local authority district (LAD).
    
    A detailed description of the SEIR state transition model used for the predictions is given in the technical note `Bayesian stochastic model-based forecasting for spatial COVID-19 risk in England' \cite{JewellTechReport2021}.
    For convenience a brief overview is given in the following.
    To simultaneously account for the variability across the LADs and their interconnectedness the model used a spatial metapopulation approach that incorporated the following explanatory variables: LAD population size; network connectivity between LADs based on commuting volume data from the 2011 Census \cite{2011CensusEngWal, 2011CensusScot, 2011CensusNI}; and traffic volume data from the United Kingdom Department for Transport (DfT) \cite{DFTtrafficVols}.
    (Population and mobility data were downloaded from the Office for National Statistics Open Geography Portal \cite{ONSgeoPortal2021}.)
    The model was fitted using a fully Bayesian approach to 12 weeks of training data.
    The COVID-19 positive tests were assumed to represent the I$\rightarrow$R transition events.
    The S$\rightarrow$E and E$\rightarrow$I transition events were therefore unobserved.
    To account for this censoring the approach taken here, based on \cite{Jewell2009}, uses a state-of-the-art MCMC algorithm within a data-augmentation framework.
    The model was fitted using the package 
    `covid19uk: Bayesian stochastic spatial modelling for COVID-19 in the UK' \cite{ChrisCovidSoftware2021} which imports from the `gemlib' repository \cite{gemlib2021}.
    A note of caution, when interpreting the estimates from the model it should be understood that the model uses data based on positive case tests consequently the estimates will be sensitive to fluctuations in testing rates.
    Ethical approval: the UKHSA COVID-19 data were supplied after anonymisation under strict data protection protocols agreed between the Lancaster University and Public Health England. The ethics of the use of these data for these purposes was agreed by Public Health England with the UK government SPI-M(O)/SAGE committees.

\begin{figure}[ht!]
    \centering
    \includegraphics[width=0.95\textwidth]{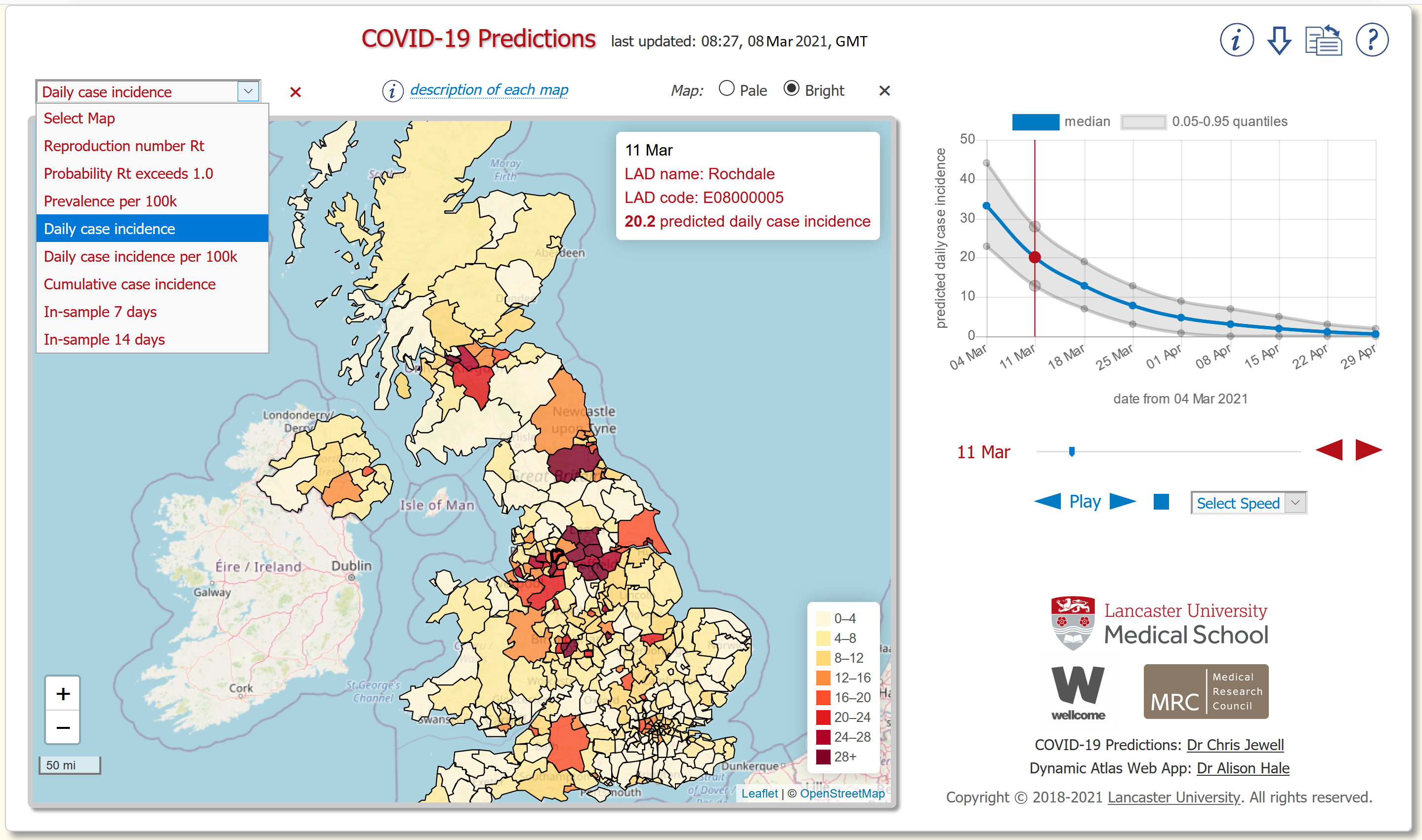}
    \caption{DHA showing COVID-19 forecasts of expected daily incidence for LADs in the UK during March and April 2021.
    The forecast depicted on the map is for 11 March 2021.
    The base map is produced by the OpenStreetMap Foundation using OpenStreetMap data under the Open Database License \cite{OpenStreetMap}.
    }
    \label{fig:covid}
    \end{figure}

    The screenshot of the DHA in Figure~\ref{fig:covid} shows forecasts relating to the COVID-19 pandemic in the United Kingdom during March and April 2021.
    For this case study the overall layout of the web app displayed in this figure incorporates the following features:
    \begin{itemize}
        \item The drop-down list (top left) used to select the prediction type includes: daily incidence and prevalence forecasts; predicted reproduction number; and insample predictions. The `Insample 7 days' (and 14 days) prediction depicts how well the prediction model fitted to the last 7 days (or 14 days) of recorded case data.
        In Figure~\ref{fig:covid} `daily case incidence' is selected.
            \item The regions on the map represent the LADs: boundary data is available from the Office for National Statistics \cite{ONSgeoPortal2021}. Each LAD is coloured coded according to the relevant prediction/forecast for that district. When an LAD is selected its corresponding time series is displayed.
        \item As indicated by the position of vertical red line on the time series in Figure~\ref{fig:covid} the map shows forecasts for 11 March 2021.
        The time series shows pointwise forecasts, along with their corresponding uncertainties, displayed at 7 day intervals from the 4 March 2021 to 29 April 2021.
        \item The inset in the top-right corner of the map gives: date of the time slice shown on the map (11 March 2021); name and code of the selected LAD (Rochdale); predicted value for the selected LAD ($20.2$).
    \end{itemize}

    The specifics of the predictions mapped in Figure~\ref{fig:covid} are as follows.
    These predictions are based on training data from 10 December 2020 to 3 March 2021.
    The expected daily incidence forecast for 11 March 2021 is shown on the map where shades of red indicate it is relatively high.
    As can be seen there are a number of areas forecast to have a relatively high level of daily case incidence around the UK: analysis of this pattern is beyond the scope of this paper for further information see for instance `COVID-19: review of disparities in risks and outcomes' \cite{PHEcovidReview2020}.
    
    Each LAD has its own time series on which the blue line is the pointwise expected value and the shaded grey area represents its corresponding uncertainty whose pointwise intervals are based on the $0.05$ and $0.95$ quantiles.
    The LAD selected in Figure~\ref{fig:covid} is Rochdale, which on the 11 March 2021 was forecast to have an absolute case incidence of $20.2$ (note that $20.2$ is the total number of expected new cases within the LAD on the given date).
    This time series for Rochdale shows that the epidemic is forecast to monotonically decrease in size, this is consistent with its predicted reproduction number of $0.50$.
    However after this forecast was made, in Rochdale there was an unexplained near doubling in daily recorded cases from the 7 March ($27$ cases) to 9 March ($55$ cases), in response subsequent model predictions/forecasts for this LAD were revised upwards. The actual number of daily recorded cases on the 11 March in Rochdale was $43$, by the beginning of April they had halved.
    This highlights the difficulty associated with making these forecasts, they arose due to the significant quantity of noise associated with the daily recorded case data and the latent complexity of the underlying spatial-temporal dynamics of COVID-19 at population-level.
    During the epidemic there were significant unobserved population-level behavioural changes as well as the emergence of new variants, both of which were influencing the case incidence.
    On the 8 March 2021 the epidemic was forecast to reduce in size in all LADs across the United Kingdom, this was reasonable given the United Kingdom was under national lockdown restrictions for the first four months of 2021.
    Daily recorded case data would later show that across the United Kingdom the epidemic continued to reduce in size from the beginning of March 2021 until at least the end of April 2021 \cite{GOVUKcovid19casedata} e.g. on 4 March 2021 there were $5,717$ daily recorded cases whereas by 30 April 2021 there were $1,692$.

\section{Discussion}

    In general a dashboard is a graphical user interface that presents an at-a-glance view of key performance indicators within a specific context, often it is hosted on web pages.
    Usually a dashboard is linked, via a pipeline, to a database so that it can be updated regularly.
    There are numerous commercial frameworks, some with free versions, for building versatile dashboard applications for example: Sisense \cite{Sisense}, Power BI \cite{MicrosoftPBI}, Tableau \cite{Tableau} and Cognos Analytics \cite{IBMCognos}.
    Typically these would be configured to display figures and tables relevant to the target audience’s requirements.
    Although the aforementioned frameworks are capable of displaying time series figures, for most purposes they would be sub-optimal or unsuitable for displaying geospatial data.
    Conversely there are many dashboards/frameworks for mapping geospatial data, these include:
    ArcGIS \cite{ArcGIS}, ArcReader \cite{arcReader}, eSpatial \cite{eSpatial} and QGIS \cite{QGIS}.
    However these are optimised for mapping geospatial data but for the most part are sub-optimal or unsuitable for conveying temporal data.
    
    Creating a dashboard which maps geospatial data alongside its corresponding temporal data generally requires a bespoke solution and developer expertise is typically required.
    For instance, Leaflet \cite{Leafletjs2018} could be configured to show a map displaying geospatial data and an image of a time series (e.g. JPG, PNG, etc.) could appear in a pop-up box when the end-user interacts with a region, or point, on the map.
    Displaying the time series graphics in a pop-up box would require a developer to write a small amount of bespoke JavaScript code and moreover the time series images would need to be created/stored in advance on the web server so they were available when the end-user interacted with the map.
    Another solution for displaying both geospatial and temporal data side-by-side would be for a developer to code a bespoke Shiny app \cite{shinyForR}, perhaps utilising flexdashboard \cite{flexdashboard}, again making use of an open source geographic information system library such as Leaflet.
    Although Shiny and Dash are a highly flexible frameworks for building data-driven web applications such as dashboards they have the drawback of requiring a specially configured web server and furthermore the browser-server communications are not lightweight.
    For instance Shiny is potentially costly in terms of host server resources which could be problematic when a number of web browsers simultaneously interact with Shiny web applications hosted on a single server, for further information see Shiny Server Administrators guide \cite{shinyAdmin}.
    These limitations were the main reasons the DHA was not developed using frameworks such as Dash and Shiny.
    However in the future as these frameworks are developed/refined perhaps they should be reconsidered in terms of creating a dynamic interactive geospatial-temporal dashboard such as the DHA.
    
    The main strength of the DHA is that it is a customisable dashboard optimised for mapping geospatial-temporal data along with its corresponding uncertainty.
    The UI/UX were a central design consideration which lead to the spatial and temporal data being displayed side-by-side, this gives a better sense of the dynamics underlying the data than would otherwise be possible if these data were displayed separately.
    The facility to interact with the maps and time series plots gives each end-user the flexibility to select the spatial and temporal slices which is of most interest to themselves.
    The option to sequentially cycle through maps at successive time slices allows the end-user to gain a sense of both the temporal and spatial dynamics.
    The DHA is straightforward for an end-user, e.g. a policy-marker, to use which should allow them to concentrate on interpreting the data.
    In addition given the web app is open-source a future developer could build upon it for their own purposes, for example the app's web page could easily be customised to fit a corporate image or its functionality could be extended to convey additional data on the time series etc.
    
    The DHA can be implemented as stand-alone web app on a web server, in which case its corresponding configuration app \cite{ShinyDHA} (described in Section \ref{sec:ShinyWebApp}) is one method of generating the required configuration and adapted-GeoJSON files.
    Furthermore this configuration app can be used directly by an end-user to display their geospatial-temporal data however this has the limitation that the data is not stored beyond the end-users' web browser session.
    Overall this configuration app assists the end-user/developer but future versions could look at ways to improve its functionality.
    For example, the configuration app could create a clone of the fully working DHA complete with the developers' data, this clone could be made available for download.
    The clone could then be copied onto the developer's web server as a fully functioning DHA thereby requiring no further configuration.
    Alternatively a future version of the DHA could be written purely using JavaScript with an improved architecture for configuring the GeoJSON and configuration files: for instance it could more closely follow the configuration methods commonly used by other JavaScript libraries such as Leaflet \cite{Leafletjs2018} and Chart.js \cite{Chartjs2018}.
    
    Another strength of the DHA is that it is relatively lightweight with regard to browser-server communications partly because it preloads everything, including all data, at the point it is first called by the web browser.
    Arguably preloading everything is also a limitation in so much as the web app plus its data is constrained by the amount of available web browser memory.
    The full web app including all open source JavaScript libraries is approximately 1MB, this excludes the GeoJSON data.
    The canine gastroenteric case study had approximately 0.4MB of GeoJSON data whereas the COVID-19 case study had about 2.4MB.
    A pragmatic, albeit less than ideal, workaround would be to display less data, for example fewer time points.
    In the future this web browser memory issue could be addressed by modifying the JavaScript code behind the DHA so that it loads data using streaming, although this would require more calls between the web browser and host web server.
    An alternative approach to optimising web browser memory would be to consider methods for minimising the size of geospatial-temporal data files e.g. using data compression techniques.
    Future versions of the web app need to give careful consideration to the balance between loading data, data compression and browser-server communications.

    A difficulty with dashboards which require regular updates is often creating a reliable data-feed from the source data to the dashboard data i.e. creating a robust pipeline.
    This is particularly problematic if the data is sourced from multiple external locations and/or needs a considerable amount of preprocessing prior to being transferred to the dashboard.
    These difficulties occur in both case studies given in this paper, note that each has its own pipeline.
    Specifically, both studies use multiple data sources which are fed into a statistical model that uses an MCMC algorithm, the output of which needs to be distilled into a suitable format for the DHA.
    In this context it would be of future interest to look at methods for generalising pipeline architecture.
    This should include using storage methods for keeping each set of MCMC samples from each pipeline run, this would allow historic predictions to be examined retrospectively e.g. on the DHA.
    Although it would have been occasionally useful neither of the case studies given in this paper had the facility to easily view historic predictions: note that historic predictions could be displayed through bespoke methods.
    Furthermore future pipelines should also include MCMC diagnostics which an administrator checks each time the MCMC algorithm has run.
    For the canine gastroenteric outbreak MCMC diagnostics, including figures and metrics, were hosted on a web page.
    Ideally such a pipeline would be fully automated but this would only be possible if sufficiently robust diagnostic methods could be employed to halt it in the event that the MCMC algorithm fully or partially failed e.g. due to poor mixing of the chain.

    \textbf{Summary}: The DHA was designed to be an interactive visualisation tool for displaying geospatial-temporal data along with its corresponding uncertainty.
    An acknowledgement of uncertainty is important as neither empirical data or model based predictions are exact.
    In contrast to many health maps which are static the DHA displays geospatial data on a map beside temporal data on a time series, therefore it possible to scroll through time to see the spatial evolution of data on the map and also examine the temporal evolution of each region on a time series.
    The web app is open-source and predominantly coded in JavaScript, it makes extensive use of two open-source libraries Leaflet and Chart.js for displaying the maps and time series respectively.
    Two case studies which used the DHA were presented in the paper: both had their own pipelines running from source data, to predictions based on Bayesian inference, through to predictions formatted for the web app.
    The first case study related to an outbreak of canine gastroenteric disease in the United Kingdom during early 2020, at each veterinary premise the DHA reported the probability of an unusually high case incidence.
    In the second case study the DHA displayed, for each local authority district in the United Kingdom during early 2021, COVID-19 predictions including case incidence and prevalence forecasts.
    Both case studies showed the DHA to be an effective tool for communicating geospatial-temporal data with a complex structure.
    Consequently, by being able to appreciate simultaneously the spatial and temporal dynamics end-users, such as policy-makers, should be able to gain good insights into these data/predictions, in turn this has the potential to enhance effective decision-making.

\section*{Data availability}

Canine gastroenteric disease outbreak: the datasets generated and/or analysed are not publicly available due to issues of companion animal owner confidentiality, but are available on reasonable request from the SAVSNET Data Access and Publication Panel (savsnet@liverpool.ac.uk) for researchers who meet the criteria for access to confidential data.

COVID-19 outbreak: The COVID-19 data were obtained from the UK Health Security Agency. These data contain confidential information, with public data deposition non-permissible for socioeconomic reasons. Requests for this data should be made to the UK Health Security Agency.

\section*{Competing interests}
    The authors declare no competing interests.

\section*{Author contributions}
    The Dynamic Health Atlas web app was conceived by A.C.H., P.J.D.
    This web app was designed and programmed by A.C.H.
    Canine gastroenteric disease outbreak contributors: C.A., A.C.H., C.P.J., P.-J.M.N., G.L.P., B.R., A.D.R.
    COVID-19 outbreak contributors: C.P.J., A.C.H., B.R.  
    The manuscript was drafted by A.C.H.
    The manuscript was revised critically for important intellectual content by all authors.
    All authors gave final approval for publication.

\section*{Acknowledgments}
    A.C.H. was funded by Medical Research Council `Map-based Visualisation and Statistical Inference with Dynamic Health Data' (MR/N015266/1) and the UKRI through the JUNIPER Consortium (MR/V038613/1).

    C.A., P.-J.M.N, G.L.P., and A.D.R were in part by the Dogs Trust as part of SAVSNET-Agile, and by the Biotechnology and Biological Sciences Research Council, and previously by the British Small Animal Veterinary Association.
    
    C.P.J. was funded by the Wellcome Trust `GEM: translational software for outbreak analysis' and the UKRI through the JUNIPER Consortium (MR/V038613/1).
    
    The authors are very grateful to Professor Niels Peek (Division of Informatics, Imaging \& Data Sciences at The University of Manchester, UK) for his invaluable contribution to the early development of the Dynamic Health Atlas.

    SAVSNET is extremely grateful to the veterinary practices across the UK who supply their data for free, and without whom these datasets would not be available.
    
    The authors would like to thank The High End Computing facility at Lancaster University for providing the facilities required for the COVID-19 model predictions.
    
    The views expressed in this paper are those of the authors and not necessarily those of their respective funders or institutions.

\bibliography{refs}{}

\begin{thebibliography}{10}

\bibitem{Reinsel2018}
D.~Reinsel, J.~Gantz, and J.~Rydning, ``Data age 2025: The digitization of the
  world from edge to core,'' {\em International Data Corporation, Framingham,
  US}, no.~US44413318, 2018.

\bibitem{NatMed2020}
{Focus: editorial}, ``Big hopes for big data,'' {\em Nature Medicine}, vol.~81,
  no.~26, 2020.

\bibitem{Tomines2013}
A.~Tomines, H.~Readhead, A.~Readhead, and S.~M. Teutsch, ``Applications of
  electronic health information in public health: Uses, opportunities and
  barriers,'' {\em eGEMs (Generating Evidence \& Methods to improve patient
  outcomes)}, vol.~1, no.~2, 2013.

\bibitem{WHOmaps}
World Health Organization (WHO), {\em The Global Health Observatory, Map
  Gallery}.
\newblock Available at \url{https://www.who.int/data/gho/map-gallery/}.

\bibitem{IHMEmaps}
The Institute for Health Metrics and Evaluation (IHME), {\em GBD (Global Burden
  of Disease) Data Visualizations}.
\newblock Available at
  \url{http://www.healthdata.org/gbd/data-visualizations/}.

\bibitem{SAHSUmaps}
UK Small Area Health Statistics Unit (SAHSU), {\em The Environment and Health
  Atlas for England and Wales}.
\newblock Available at \url{https://www.envhealthatlas.co.uk/}.

\bibitem{HealthMapBCH}
Boston Children's Hospital, Computational Epidemiology Lab, Boston, {\em
  HealthMap}.
\newblock Available at \url{https://healthmap.org/en/}.

\bibitem{covidCSSEJHU}
Center for Systems Science and Engineering (CSSE), Johns Hopkins University,
  {\em COVID-19 Dashboard}.
\newblock Available at
  \url{https://gisanddata.maps.arcgis.com/apps/opsdashboard/index.html#/bda7594740fd40299423467b48e9ecf6}.

\bibitem{covidBIUV}
Biocomplexity Institute, University of Virginia, {\em COVID-19 Surveillance
  Dashboard}.
\newblock Available at
  \url{https://nssac.bii.virginia.edu/covid-19/dashboard/}.

\bibitem{dashPiFr}
{\em {D}ash: {P}ython framework for building web analytic applications}.
\newblock Available at \url{https://dash.plotly.com/};
  \url{https://github.com/plotly/dash}.

\bibitem{Shiny}
{RStudio, Inc}, {\em Shiny from RStudio}, 2021.
\newblock Available at \url{https://shiny.rstudio.com/}.

\bibitem{Leafletjs2018}
{\em Leaflet: an open-source JavaScript library for mobile-friendly interactive
  maps. Version 1.3.1}.
\newblock Available at \url{https://leafletjs.com/}.

\bibitem{Chartjs2018}
{\em Chart.js: Simple yet flexible JavaScript charting for designers \&
  developers. Version: 2.7.1}.
\newblock Available at \url{https://www.chartjs.org}.

\bibitem{Butler2016}
H.~Butler, M.~Daly, A.~Doyle, S.~Gillies, S.~Hagen, and T.~Schaub, ``The
  {GeoJSON} format,'' {\em RFC 7946}, 2016.

\bibitem{DHAtutorial}
A.~C. Hale, {\em Dynamic Health Atlas web app - tutorial}, 2021.
\newblock Available at \url{https://achale.gitlab.io/dynamicatlastutorial/}.

\bibitem{DHAcode}
A.~C. Hale, {\em Dynamic Health Atlas web app - source code}, 2021.
\newblock Available at \url{https://gitlab.com/achale/dynamicatlas}.

\bibitem{DHAdemo}
A.~C. Hale, {\em Dynamic Health Atlas web app - demo}, 2021.
\newblock Available at \url{https://achale.gitlab.io/dynamicatlas/}.

\bibitem{shinyForR}
W.~Chang, J.~Cheng, J.~J. Allaire, C.~Sievert, B.~Schloerke, Y.~Xie, J.~Allen,
  J.~McPherson, A.~Dipert, and B.~Borges, {\em shiny: Web Application Framework
  for {R}}.
\newblock Available at \url{https://CRAN.R-project.org/package=shiny}.

\bibitem{Rcran2021}
{R Core Team}, {\em R: A Language and Environment for Statistical Computing}.
\newblock R Foundation for Statistical Computing, Vienna, Austria, 2021.

\bibitem{ESRI1998}
ESRI, {\em ESRI Shapefile Technical Description}, 1998.
\newblock Available at \url{https://support.esri.com/en/white-paper/279}.

\bibitem{ShinyDHA}
A.~C. Hale, {\em Shiny app for Dynamic Health Atlas}.
\newblock Available at
  \url{http://fhm-chicas-apps.lancs.ac.uk/shiny/users/haleac/healthatlas/}.

\bibitem{Fernando2015}
F.~S{\'a}nchez-Vizca{\'i}no, P.~H. Jones, T.~Menacere, B.~Heayns, M.~Wardeh,
  J.~Newman, A.~D. Radford, S.~Dawson, R.~Gaskell, P.-J.~M. Noble, S.~Everitt,
  M.~J. Day, and K.~McConnell, ``Small animal disease surveillance,'' {\em
  Veterinary Record}, vol.~177, no.~23, pp.~591--594, 2015.

\bibitem{Fernando2017}
F.~S{\'a}nchez-Vizca{\'i}no, P.-J.~M. Noble, P.~H. Jones, T.~Menacere,
  I.~Buchan, S.~Reynolds, S.~Dawson, R.~M. Gaskell, S.~Everitt, and A.~D.
  Radford, ``Demographics of dogs, cats, and rabbits attending veterinary
  practices in {Great Britain} as recorded in their electronic health
  records,'' {\em BMC Veterinary Researchs}, vol.~13, no.~218, 2017.

\bibitem{ONeill2014}
D.~G. O{\textquoteright}Neill, D.~B. Church, P.~D. McGreevy, P.~C. Thomson, and
  D.~C. Brodbelt, ``Prevalence of disorders recorded in dogs attending
  primary-care veterinary practices in {E}ngland,'' {\em PLOS ONE}, vol.~9,
  no.~3, pp.~1--16, 2014.

\bibitem{Radford2021}
A.~D. Radford, D.~A. Singleton, C.~Jewell, C.~Appleton, B.~Rowlingson, A.~C.
  Hale, C.~T. Cuartero, R.~Newton, F.~S{\'a}nchez-Vizca{\'i}no, D.~Greenberg,
  B.~Brant, E.~G. Bentley, J.~P. Stewart, S.~Smith, S.~Haldenby, P.-J.~M.
  Noble, and G.~L. Pinchbeck, ``Outbreak of severe vomiting in dogs associated
  with a canine enteric coronavirus, {United Kingdom},'' {\em Emerging
  Infectious Diseases}, vol.~27, no.~2, 2021.

\bibitem{caramellar2017}
B.~Rowlingson, E.~Giorgi, and A.~C. Hale, {\em Conditional Auto-regressive
  Space-Time Model (caramellar)}, 2021.
\newblock Available at
  \url{https://github.com/barryrowlingson/caramellar/tree/master}.

\bibitem{Hale2019}
A.~C. Hale, F.~S{\'a}nchez-Vizca{\'i}no, B.~Rowlingson, A.~D. Radford,
  E.~Giorgi, S.~J. O{\textquoteright}Brien, and P.~J. Diggle, ``A real-time
  spatio-temporal syndromic surveillance system with application to small
  companion animals,'' {\em Scientific Reports}, vol.~9, 2019.

\bibitem{ONSgeoPortal2021}
Office for National Statistics, United Kingdom, {\em Open Geography portal},
  2021.
\newblock Available at \url{https://geoportal.statistics.gov.uk/}.

\bibitem{doogal2020}
C.~Bell, {\em doogal.co.uk}, 2020.
\newblock Available at \url{https://www.doogal.co.uk/}.

\bibitem{OpenStreetMap}
OpenStreetMap Foundation, {\em OpenStreetMap data licensed under the Open Data
  Commons Open Database License}.
\newblock Available at \url{https://www.openstreetmap.org/};
  \url{https://www.openstreetmap.org/copyright}.

\bibitem{DHSCPillars2021}
Department of Health and Social Care, London, UK, {\em Coronavirus
  ({COVID-19}): testing data methodology}.
\newblock Available at
  \url{https://www.gov.uk/government/publications/coronavirus-covid-19-testing-data-methodology}.

\bibitem{GOVUKcovid19casedata}
Public Health England, London, {\em GOV.UK: Coronavirus (COVID-19) in the UK -
  The official UK Government website for data and insights on Coronavirus
  (COVID-19)}, 2021.
\newblock Available at \url{https://coronavirus.data.gov.uk/};
  \url{https://coronavirus.data.gov.uk/details/cases}.

\bibitem{JewellTechReport2021}
C.~Jewell, J.~Read, G.~Roberts, B.~Rowlingon, and C.~Suter, ``Bayesian
  stochastic model-based forecasting for spatial {COVID}-19 risk in
  {E}ngland,'' tech. rep., Lancaster Medical School, Lancaster University;
  Department of Statistics, University of Warwick; Google Research, New York, 4
  March 2021.
\newblock Available at
  \url{https://chicas-covid19.gitlab.io/bayesstm/docs/doc_lancs_space_model_concept.pdf}.

\bibitem{2011CensusEngWal}
Office for National Statistics, England, UK, {\em 2011 Census: England and
  Wales}.
\newblock Available at \url{https://www.ons.gov.uk/census/2011census}.

\bibitem{2011CensusScot}
National Records of Scotland, Scotland, UK, {\em 2011 Census: Scotland}.
\newblock Available at \url{https://www.scotlandscensus.gov.uk/census-results}.

\bibitem{2011CensusNI}
Northern Ireland Statistics and Research Agency, Northern Ireland, {\em 2011
  Census: Northern Ireland}.
\newblock Available at
  \url{https://www.nisra.gov.uk/statistics/2011-census/results}.

\bibitem{DFTtrafficVols}
Department for Transport, London, {\em Road traffic statistics}.
\newblock Available at \url{https://roadtraffic.dft.gov.uk/};
  \url{https://www.gov.uk/government/collections/road-traffic-statistics}.

\bibitem{Jewell2009}
C.~P. Jewell, M.~J. Keeling, and G.~O. Roberts, ``Predicting undetected
  infections during the 2007 foot-and-mouth disease outbreak,'' {\em Journal of
  The Royal Society Interface}, vol.~6, no.~41, pp.~1145--1151, 2009.

\bibitem{ChrisCovidSoftware2021}
C.~P. Jewell, {\em covid19uk: Bayesian stochastic spatial modelling for
  COVID-19 in the UK}, 2021.
\newblock Available at \url{https://github.com/chrism0dwk/covid19uk}.

\bibitem{gemlib2021}
C.~P. Jewell and A.~C. Hale, {\em gemlib: a scientific compute library build
  for epidemic analysis}, 2021.
\newblock Available at \url{http://fhm-chicas-code.lancs.ac.uk/GEM/gemlib}.

\bibitem{PHEcovidReview2020}
Public Health England, London, {\em COVID-19: review of disparities in risks
  and outcomes}, 11 August 2020.
\newblock Available at
  \url{https://www.gov.uk/government/publications/covid-19-review-of-disparities-in-risks-and-outcomes}.

\bibitem{Sisense}
Sisense, {\em Power to the analytics builders}.
\newblock Available at \url{https://www.sisense.com/en-gb/}.

\bibitem{MicrosoftPBI}
Microsoft, {\em Power BI}.
\newblock Available at \url{https://powerbi.microsoft.com/en-us/}.

\bibitem{Tableau}
Tableau Software, {\em Tableau}.
\newblock Available at \url{https://www.tableau.com/en-gb/}.

\bibitem{IBMCognos}
IMB, {\em Cognos Analytics}.
\newblock Available at
  \url{https://www.ibm.com/uk-en/products/cognos-analytics/}.

\bibitem{ArcGIS}
Esri, {\em ArcGIS}.
\newblock Available at \url{https://www.arcgis.com/}.

\bibitem{arcReader}
Esri, {\em ArcReader}.
\newblock Available at
  \url{https://www.esri.com/en-us/arcgis/products/arcreader/}.

\bibitem{eSpatial}
eSpatial, {\em eSpatial}.
\newblock Available at \url{https://www.espatial.com/}.

\bibitem{QGIS}
Open Source Geospatial Foundation, {\em QGIS Geographic Information System}.
\newblock Available at \url{https://qgis.org/}.

\bibitem{flexdashboard}
{\em flexdashboard: publish a group of related data visualizations as a
  dashboard using {R Markdown}}.
\newblock Available at \url{http://rmarkdown.rstudio.com/flexdashboard};
  \url{https://CRAN.R-project.org/package=flexdashboard }.

\bibitem{shinyAdmin}
RStudio, {\em Shiny Server Open Source v1.4.0 Administrator's Guide}.
\newblock Available at \url{https://rstudio.github.io/shiny-server/os/latest/}.

\end{thebibliography}
\bibliographystyle{ieeetr}

\bigskip

\appendix
\section{Appendix}

Example of the GeoJSON format adapted for geospatial-temporal data:
\begin{Verbatim}[fontsize=\footnotesize]
{
    "type": "FeatureCollection",
    "crs": {
                "type": "name",
                "properties": {"name":"urn:ogc:def:crs:OGC:1.3:CRS84"}
            },
    "features":[
        {...},
        {
            "type": "Feature",
            "properties":{
                "area": "name",
                "0": 12,
                "1": 15,
                "2": 10,
                ...
                "0.L": 7,
                "1.L": 9,
                ...
                "0.U": 19,
                "1.U": 22,
                ...
            },
            "geometry": {
                "type": "MultiPolygon",
                "coordinates": [
                    [ [ [-1.2,54.7], [...,...], [-1.2,54.7] ] ]
                ]
            }
        },
        {...}
    ]
}
\end{Verbatim}

\bigskip

\bigskip

\bigskip

\begin{figure}[htb!]
\centering
    \includegraphics[width=0.95\textwidth]{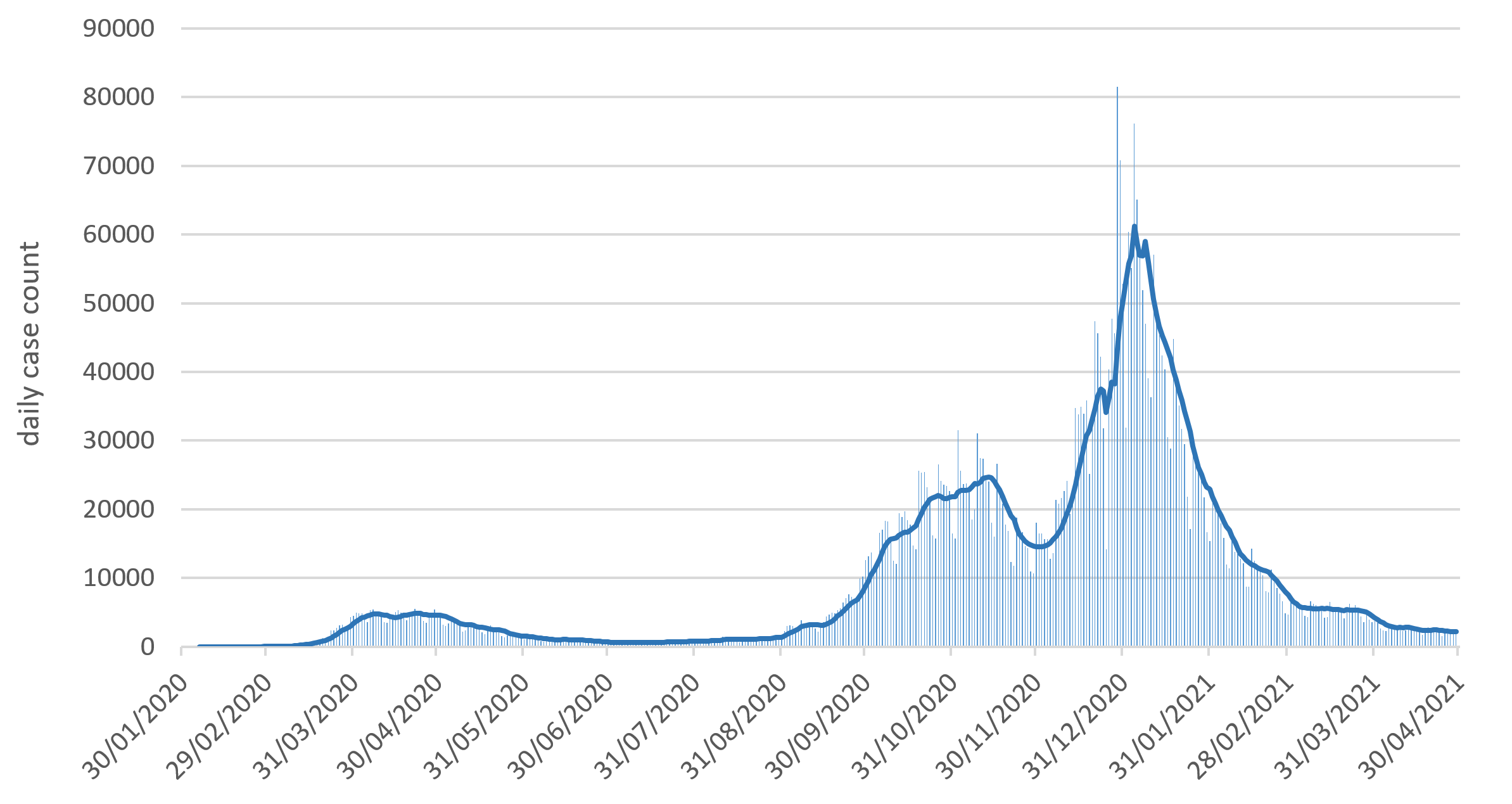}
    \caption{Daily recorded UK COVID-19 cases from 30/01/2020 to 30/04/2021.
    The solid blue line is the 7-day moving average.
    Data source: Public Health England \cite{GOVUKcovid19casedata}
    }
\label{fig:PHEcases}
\end{figure}

\end{document}